\documentclass[
    aps,
    prd,
    reprint,
    groupedaddress,
    superscriptaddress,
    nolongbibliography,
    nobibnotes,
    nofootinbib,
]{revtex4-2}

\usepackage[dvipsnames, usenames]{xcolor}
\definecolor{linkcolor}{rgb}{0.0,0.3,0.5}

\usepackage[
colorlinks=true,
linkcolor=linkcolor,
citecolor=linkcolor,
filecolor=linkcolor,
urlcolor=linkcolor,
pdfusetitle,
]{hyperref}

\usepackage{amssymb}
\usepackage{amsmath}
\usepackage{physics}
\usepackage{orcidlink}
\usepackage{xspace}
\usepackage{multirow}
\usepackage{booktabs}
\usepackage[absolute,overlay]{textpos}

\newcommand{\comment}[1]{}

\NewDocumentCommand{\spinformation}{O{1.3}}{%
 \NoCaseChange{\raisebox{-0.1ex}{%
   \includegraphics[height=#1\fontcharht\font`X]{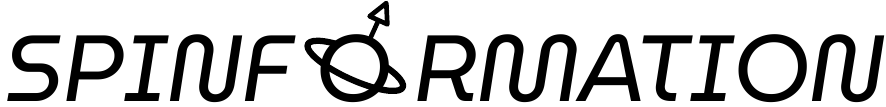}}}\xspace
}

\newcommand{\chieff}{\ensuremath{\chi_\mathrm{eff}}\xspace}
\newcommand{\chip}{\ensuremath{\chi_\mathrm{p}}\xspace}
\newcommand{\chipgen}{\ensuremath{\chi_\mathrm{p,gen}}\xspace}
\newcommand{\kleff}{KL\textsubscript{eff}\xspace}
\newcommand{\klp}{KL\textsubscript{p}\xspace}
\newcommand{\cz}[1]{\ensuremath\chi_{z,#1}\xspace}
\newcommand{\cperp}[1]{\ensuremath\chi_{\perp,#1}\xspace}

\newcommand{\ligo}{\affiliation{LIGO Laboratory, Massachusetts Institute of Technology, Cambridge, MA 02139, USA}}
\newcommand{\mki}{\affiliation{Kavli Institute for Astrophysics and Space Research, Massachusetts Institute of Technology, Cambridge, MA 02139, USA}}
\renewcommand{\mit}{\affiliation{Department of Physics, Massachusetts Institute of Technology, Cambridge, MA 02139, USA}}
\newcommand{\flatiron}{\affiliation{Center for Computational Astrophysics, Flatiron Institute, 162 5th Ave, New York, NY
10010}}
\newcommand{\milan}{\affiliation{Dipartimento di Fisica ``G. Occhialini'', Universit\'a degli Studi di Milano-Bicocca, Piazza della Scienza 3, 20126 Milano, Italy}}
\newcommand{\infn}{\affiliation{INFN, Sezione di Milano-Bicocca, Piazza della Scienza 3, 20126 Milano, Italy}}

\begin{document}

\begin{textblock*}{4cm}(0.5pt,5pt)
    \includegraphics[width=4cm]{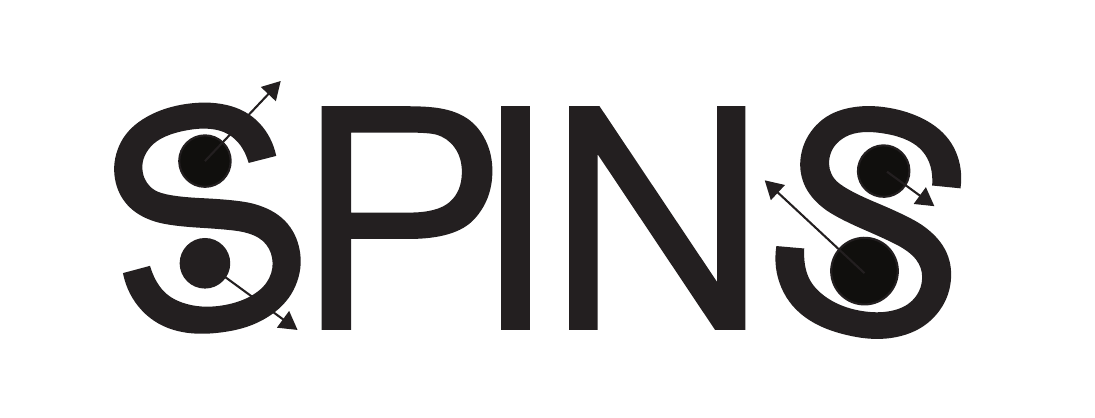}
\end{textblock*}

\title{\texorpdfstring{\spinformation}{Spinformation}: do gravitational-wave events carry\\information beyond effective spins?}

\author{Cailin Plunkett\,\orcidlink{0000-0002-1144-6708}}\email{caplunk@mit.edu}
\ligo\mki\mit

\author{Jack Heinzel\,\orcidlink{0000-0002-5794-821X}}
\ligo\mki\mit\flatiron

\author{Noah E. Wolfe,\,\orcidlink{0000-0003-2540-3845}}
\ligo\mki\mit

\author{Davide Gerosa$\,$\orcidlink{0000-0002-0933-3579}}
\milan \infn

\author{Salvatore Vitale\,\orcidlink{0000-0003-2700-0767}}
\ligo\mki\mit

\collaboration{Society of Physicists Interested in Non-aligned Spins, SPINS}
\thanks{\url{www.sites.mit.edu/spins}}

\date{\today}

\begin{abstract}

While the spin vectors of binary black holes offer insight into the astrophysical formation channels of these systems, individual spins are less precisely measured through gravitational-wave observations than the effective inspiral spin \chieff and the effective precession spin \chip.
Reducing to effective parameters discards information in the data, but how much information is lost has yet to be demonstrated. Here we show how to quantify whether individual mergers carry information about component spins beyond the constraints implied by effective parameters (a.k.a. \spinformation) and apply the technique to GWTC-5. Thresholding on information loss $\gtrsim 0.25$ bits, over $80\%$ of gravitational-wave events in GWTC-5 have component spin posteriors described by \chieff alone. Only $\sim3\%$ of gravitational-wave events have substantial spin information not explained by \chieff and \chip. The catalog's component spin information is effectively two-dimensional in \chieff and \chip. Using simulated signals, we find the noise realization can create or destroy over a bit of beyond-\chieff information: the structure in real data may therefore be driven by noise. At current sensitivity, component spins provide interpretive---rather than informational---advantage over effective parameters.

\end{abstract}

\maketitle

\section{Introduction}

The spin vectors of binary black holes (BBHs) are useful discriminators between the astrophysical environments that may produce merging compact binaries~\cite[see, e.g.,][for reviews]{Mandel:2018hfr, Mapelli:2021taw}. However, spin effects in gravitational-wave (GW) signals are predominantly driven by degenerate combinations of spin components known as effective parameters~\cite{Racine:2008qv, Ajith:2009bn, Ajith:2011ec, Purrer:2013ojf, Schmidt:2014iyl}.

Spin parameters first appear in the GW phase evolution at 1.5 post-Newtonian order as spin-orbit coupling terms, often parameterized using the effective spin~\cite{Racine:2008qv, Ajith:2009bn}:
\begin{equation}\label{eq:chieff} 
    \chieff = \frac{a_1\cos\tau_1+q\,a_2\cos\tau_2}{1+q},
\end{equation} 
where subscripts 1 (2) indicate the more (less) massive black hole (BH), $a$ is the spin magnitude, $\tau$ the polar angle (tilt) with respect to the orbital angular momentum at a chosen reference frequency, and $q\leq1$ the binary mass ratio. The effective spin is a constant of motion to second post-Newtonian order~\cite{Racine:2008qv, Blanchet:2013haa,Gerosa:2015tea}.
Better-measured than the constituent component spins, its measurability is also well-characterized across parameter space~\cite{Purrer:2013ojf, Vitale:2016avz, Ng:2018neg, Chatziioannou:2018wqx, Shaik:2019dym, Biscoveanu:2021nvg,Wolfe:2023yuu}. Informative constraints on the \chieff of individual GW events have occurred since early observations~\cite{LIGOScientific:2018mvr}.

Spinning BH in a binary may also induce precession of the orbital angular momentum, the magnitude of which is commonly quantified by the effective precessing spin~\cite{Schmidt:2014iyl}:
\begin{equation}\label{eq:chip}
    \chip = \max\left(a_1\sin\tau_1,\,q\frac{3+4q}{4+3q}a_2\sin\tau_2\right).
\end{equation} While widely employed, \chip is not conserved during inspiral and has inconsistencies that have motivated corrected parameterizations~\cite{Gerosa:2020aiw, DeRenzis:2022vsj} and a two-dimensional alternative~\cite{Thomas:2020uqj}.
Unambiguous measurements of nonzero \chip are sought as both confirmation of general-relativistic dynamics and signatures of formation channels~\cite{LIGOScientific:2018mvr, Baibhav:2021qzw, Payne:2024ywe, Biscoveanu:2026ikx}, but have remained elusive except in a handful of cases~\cite{LIGOScientific:2020iuh, LIGOScientific:2025rsn, LIGOScientific:2025brd, LIGOScientific:2026wfs}, as precession is typically weakly constrained~\cite{Fairhurst:2019srr, Green:2020ptm}. Evidence for precession may further be impacted by waveform systematics and data-quality issues~\cite{Ossokine:2020kjp, Pratten:2020ceb, Hoy:2021dqg, Payne:2022spz, Udall:2024ovp, Macas:2023wiw, Ray:2025rtt, Udall:2025bts, LIGOScientific:2025rsn, Bini:2026kwz}, as can orbital eccentricity masquerade as precession~\cite{Gayathri:2020coq, Romero-Shaw:2022fbf, Fumagalli:2023hde, Jan:2025zcm, Tibrewal:2026jci}. The pair of \chieff and \chip is often used to summarize the spin behavior of observed BBHs.

When analyzing the population distribution of BBH spins, inference can be performed in effective parameters or component spins. 
These approaches trade off between physical relevance and measurability. Analyses in effective parameter space benefit from the improved constraints on \chieff compared to component spins, while its distribution (and joint with \chip) can still probe formation channels~\cite[e.g.,][]{Liu:2018nrf, Su:2020vda, Callister:2021fpo, Fishbach:2022lzq, Callister:2022qwb, Mould:2022ccw, Antonini:2024het, Li:2025rhu, Plunkett:2026pxt, Padhyegurjar:2026scg}. Component spin models, in turn, maintain direct astrophysical interpretation~\cite[e.g.,][]{Galaudage:2021rkt, Mould:2022xeu, Callister:2022qwb, Li:2023yyt, Pierra:2024fbl, Adamcewicz:2025phm, Hussain:2026pfm}. However, poor per-event measurements, like those of component spins, can produce population results with spurious or Poisson-noise-driven features, biases that standard checks fail to identify, and model dependence~\cite{Hoy:2021rfv, Vitale:2022dpa, Miller:2024sui,  Vitale:2025lms, Miller:2026buq, Wolfe:2026meb}. Individual-event~\cite{Kobayashi:2026lac, Prasad:2026idz} and population~\cite{Wolfe:2026pjn} priors also affect population spin inference in both effective and component space; in turn, population-informed spin parameter estimates are typically less extreme than fiducial posteriors and are necessary to both define and interpret exceptional events~\cite{Mould:2026nle}.

While there may not exist a universally appropriate modeling scheme, it is necessary to test the assumptions that underlie the nominal advantages of each approach. We assess the extent to which effective parameters are a lossy compression of spin information in GW data by quantifying whether the spin magnitude and tilt posteriors\footnote{As \chieff and \chip are oblivious to azimuthal angles (although \chip should incorporate azimuthal separation~\cite{Gerosa:2020aiw}), which complete the six-dimensional spin space, we do not include these degrees of freedom in our main analysis. See Refs.~\cite{Varma:2021csh, Varma:2021xbh, Biscoveanu:2025jpc} for azimuthal measurements and future prospects for individual events and the population, as well as App.~\ref{app:chip-gen} for results using the corrected \chip.} of GW events carry information beyond the constraints implied by effective spins. This approach of \textit{prior conditioning} combines a (usually uninformative) prior with posterior information in another parameter~\cite{Gangardt:2022ltd}. It can elucidate the extent to which posterior correlations drive parameter constraints, as opposed to independent information in the data.

Prior conditioning has been employed in GW analysis to identify events for which constraints on \chip deviate from its prior conditioned on the \chieff posterior, since the two parameters are correlated~\cite[e.g.][]{LIGOScientific:2018mvr, LIGOScientific:2026wfs}. It has also been used on alternative estimators of precession dynamics derived from post-Newtonian expansions~\cite{Gangardt:2021lic, Gangardt:2022ltd}, similarly to diagnose whether constraints on precession and nutation go beyond the priors induced by $q$ and \chieff. Reference~\cite{Gangardt:2022ltd} found no events through the Third Gravitational-Wave Transient Catalog had significant measurement of subdominant spin effects.

At the population level, Ref.~\cite{Miller:2024sui} studied information beyond \chieff via simulated populations with identical \chieff but different component distributions. They found spin magnitude distributions are measurable beyond \chieff, but tilt inference can suffer insidious biases. Ref.~\cite{Wolfe:2026pjn} finds that real component spin inference extrapolates effective spin results, modulated by the population prior, and can erroneously indicate features in the tilt distribution.

Here we introduce \spinformation: how much information lies in each \textit{per-event component spin posterior} beyond the effective spin(s)? Analyzing the 259 BBHs with false alarm rate $<1\ \mathrm{yr}^{-1}$ in the Fifth Gravitational-Wave Transient Catalog (GWTC-5)~\cite{LIGOScientific:2026sit, LIGOScientific:2026wfs}, we condition component spin priors on both \chieff and $q$ as well as \chieff, $q$, \textit{and} \chip\footnote{Throughout, while we always condition on mass ratio, we frequently drop the conditional on $q$ in text to highlight the number of \textit{spin} parameters to which we reduce the data.} to identify the extent to which measurement of the four-dimensional component spin space is reducible to one (or two) effective spin parameters at current sensitivity. 

\begin{figure}
\centering
\includegraphics[width=\columnwidth]{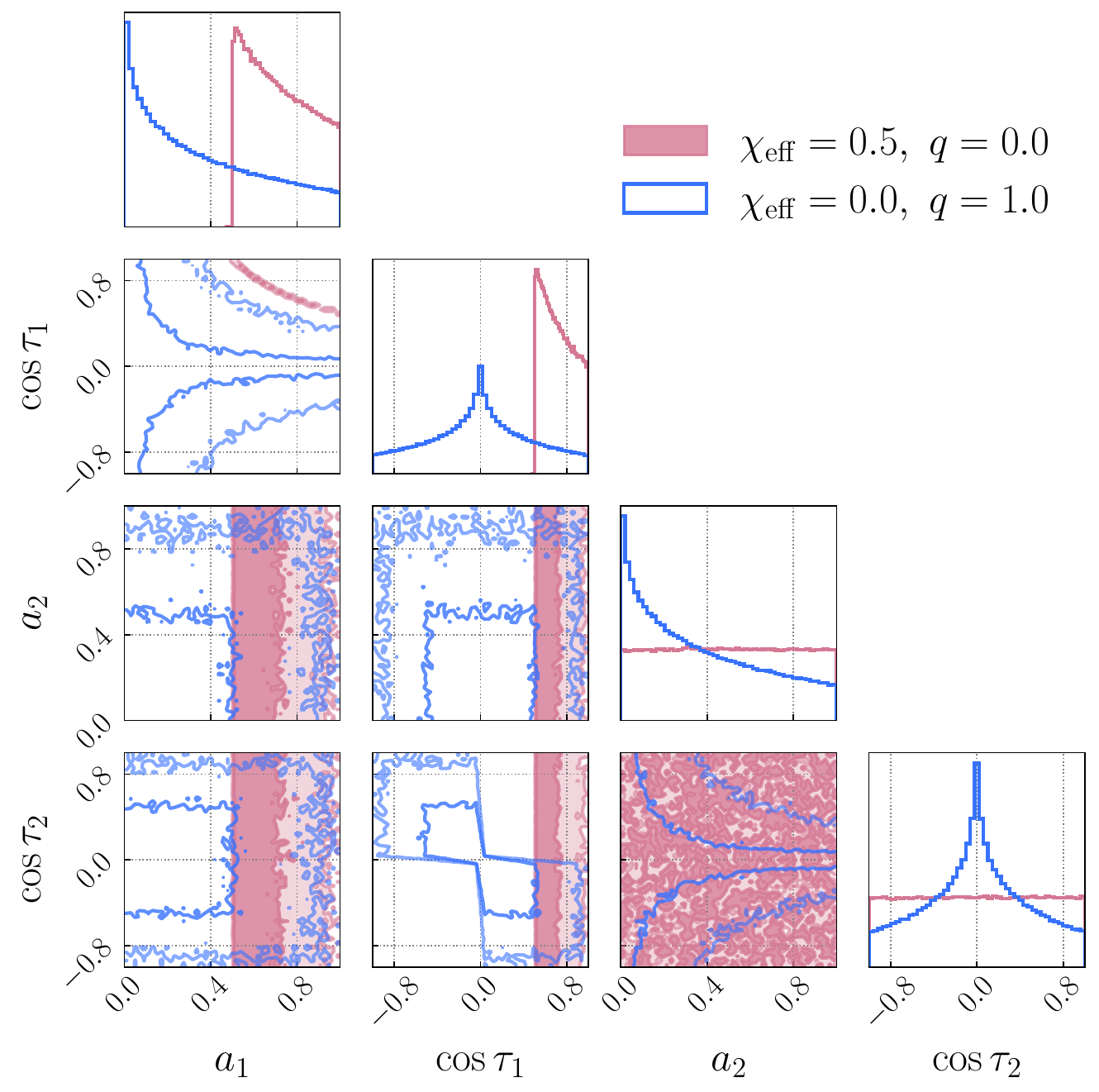}
\caption{Examples of draws from an isotropic component spin prior conditioned on ($\chieff, q$) pairs. Contours enclose the 50\% and 90\% credible regions.}
\label{fig:illustrate-cond-prior}
\end{figure}

\section{Analysis setup}\label{sec:cond-prior}

\subsection{Conditional priors}

How effective spin and mass ratio constrain component spins is encoded by the conditional prior:
\begin{equation}\label{eq:cond-prior}
p(a_1,\cos\tau_1, a_2, \cos\tau_2 \mid\chieff,q,\mathrm{iso})\,,
\end{equation} which conditions an initial isotropic prior on a $(\chieff, q)$ pair. The additional constraint \chip places on component spins is given by the density
\begin{equation}\label{eq:cond-chip-prior}
    p(a_1,\cos\tau_1, a_2, \cos\tau_2 \mid\chieff,\chip,q,\mathrm{iso})\,.
\end{equation}
Both expressions are derived in App.~\ref{app:cond-prior-chieff}.

To illustrate how \chieff and $q$ constrain component spins, we show two extremal cases of the $(\chieff, q)$-conditioned prior [Eq.~(\ref{eq:cond-prior})] in Figure~\ref{fig:illustrate-cond-prior}. In the unphysical limit of $q=0$ (pink, shown for $\chieff=0.5$), the constraint reduces to the hyperbola $\chieff = a_1\cos\tau_1$. Consequently, the secondary's properties are entirely unconstrained; the distributions of $a_2$ and $\cos\tau_2$ remain uniform across their prior range. The primary spin and tilt are exactly degenerate and are limited to magnitudes $a_1\geq|\chieff|$ and $|\cos\tau_1|\geq|\chieff|$, with \chieff and $\cos\tau_1$ of the same sign.

In the equal-mass limit with $\chieff=0$ (blue), the constraint becomes $a_1\cos\tau_1=-a_2\cos\tau_2$: the induced prior is identical on the primary and secondary. The aligned spins $a_i\cos\tau_i,\ i\in[1,2]$ must be equal and opposite, which permits both vanishing spin magnitudes and in-plane spins of any magnitude. This produces characteristic ``T'' correlations between $a_1$--$\cos\tau_1$ and $a_2$--$\cos\tau_2$. In addition, at $\chieff=0$, the tilts must be of opposite sign, creating an ``hourglass'' relating $\cos\tau_1$ to $\cos\tau_2$.

In general, the $(\chieff,q)$-conditioned prior is fully correlated. The component spins of the {primary} BH are correlated with those of the {secondary} BH, since the aligned spins must balance to produce the required \chieff. In turn, each BH's spin magnitude is correlated with its {own} tilt, since a given aligned spin sets a hyperbolic constraint. Equal masses produce the strongest correlation between the primary and secondary. Mass asymmetry upweights the primary, tightening the relation between $a_1$ and $\cos\tau_1$ while reducing the interdependence of $a_2$ and $\cos\tau_2$.

Conditioning on \chip introduces further nontrivial structure into the component spin priors. Surfaces of constant \chieff and \chip can contain two branches defined by which body dominates \chip, constraints that are asymmetric at unequal mass. As precession is a perpendicular degree of freedom to \chieff, precession measurement can break degeneracies inherent to \chieff such as that between a nonspinning and in-plane-spin BBH with $\chieff=0$.

\subsection{Quantifying \texorpdfstring{\spinformation}{Spinformation}}\label{ssec:quantifying-spinfo}

Analysts perform parameter estimation on GW data $d$ to obtain posterior samples $\vec\theta_j\sim p(\vec\theta\mid d)$ for each event. To obtain the ($\chieff,q$)-conditioned prior for each event, we take the prior on component spin parameters $\vec s = (a_1,\cos\tau_1,a_2,\cos\tau_2)$, condition on ($\chieff,q$), and marginalize over the ($\chieff,q$) posterior:
\begin{equation}
\begin{aligned}\label{eq:comp-marg-over-eff}
    \pi_{\chieff,q}(\vec s\mid d,\mathrm{iso}) &= \int \dd\chieff\, \dd q \, p(\vec s\mid \chieff,q,\mathrm{iso})\\
    &\times p(\chieff,q \mid d).
\end{aligned}
\end{equation}
In practice, for each posterior sample we draw a set of component spins from the conditioned prior, $\vec s_j^\mathrm{\ cond}\sim p(\vec s\mid \chi_{\mathrm{eff}\,j},q_j,\mathrm{iso})$. We construct analogous priors conditioned on the $(\chieff,\chip,q)$ posterior.

We use the Kullback-Leibler (KL) divergence to quantify information gain from the conditioned priors to posterior, with illustrative examples in App.~\ref{app:info}. Conditioned on $(\chieff,q)$, we define
\begin{equation}
\begin{aligned}
    \mathrm{KL_{eff}} &= \mathrm{D_{KL}}\big[p(\vec s\mid d)\parallel \pi_{\chieff,q}(\vec s\mid d,\mathrm{iso})\big] \\
    &=\int p(\vec s\mid d)\log \frac{p(\vec s\mid d)}{\pi_{\chieff,q}(\vec s\mid d,\mathrm{iso})}\dd \vec s.
\end{aligned}
\end{equation}
We repeat using the prior conditioned on $(\chieff,\chip,q)$ and call this term \klp.

To estimate KL divergences from samples in four dimensions, we make use of the fact that KL is the expected value of the log density ratio and train binary classifiers for density ratio estimation~\cite{Sugiyama_Suzuki_Kanamori_2012, Cranmer:2015bka}, detailed in App.~\ref{app:classifier}. We set a heuristic limit of \kleff or \klp $>0.25$ bits as evidence of significant information beyond the effective parameter(s) (App.~\ref{app:info}). Given our findings regarding the scatter due to noise in Sec.~\ref{ssec:noise}, we additionally report event totals under higher thresholds of \spinformation \kleff $>1$ and \klp $>0.5$ bits.

The four-dimensional KLs describe the total beyond-effective information in the joint component spin posterior. We are also interested in how this information is distributed across the component spin parameters. To extract which individual components tend to be better-measured than effective parameters require, we compute the beyond-effective information in the marginal distribution of each component $x\in\vec s$, which we call $\mathrm{KL_{eff/p}}(x)$.

\section{\texorpdfstring{\spinformation[1.75]}{Spinformation} in GWTC-5}\label{sec:results}

We analyze the beyond-effective \spinformation in GWTC-5 using the waveforms and samples specified in App.~\ref{app:samples}. We focus on BBH events in the main text but present the \spinformation for neutron-star-containing events in App.~\ref{app:nsbh}.

\begin{figure}
\centering
\includegraphics[width=\columnwidth]{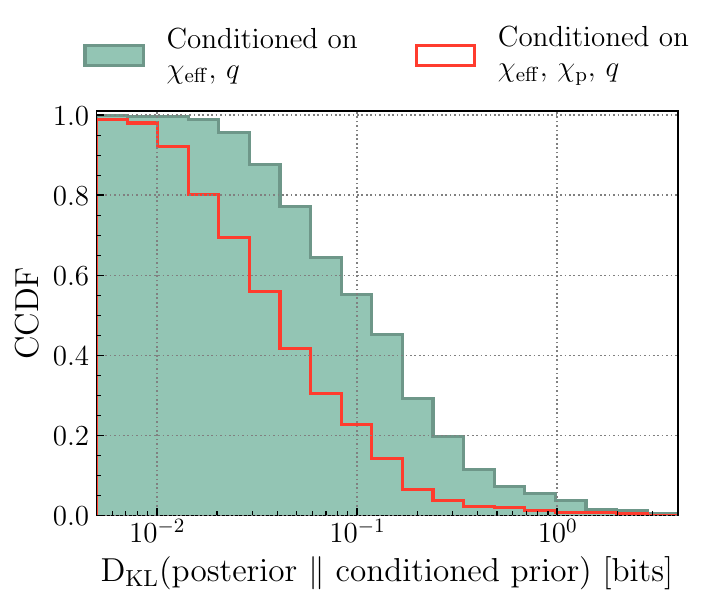}
\caption{The complementary CDF of component spin information beyond effective spin(s) in GWTC-5. The filled histogram shows information beyond \chieff, while the step red histogram provides that for information beyond both \chieff and~\chip.}
\label{fig:spinformation-distribution}
\end{figure}

 The complementary cumulative distribution functions (CDFs) of beyond-\chieff and beyond-(\chieff,\chip) information in the catalog are in Fig.~\ref{fig:spinformation-distribution}.
In GWTC-5, 48 events (18\%) have at least 0.25 bits of information beyond \chieff in some parameter; nine events have $>1$ bit. Under the more relaxed threshold, effective spin is an adequate compression of the data's component spins for over three quarters of observed BBHs.

Few events have \spinformation beyond both effective parameters. Only 10 events (3\%) have \klp $>0.25$ bits (five have \klp $>0.5$ bits). Put another way, \chieff and \chip almost fully describe the catalog's component spin constraints. The few (\chieff,\chip)-informative events and their KLs are listed in Table~\ref{tab:real-events} alongside maximum-likelihood (ML) SNR and inferred primary mass, \chieff, and \chip. As standard \chip incorrectly averages over azimuthal separation, we also condition on a generalized precession parameter~\cite{Gerosa:2020aiw} and obtain comparable constraints (App.~\ref{app:chip-gen}), indicating azimuthal angle measurements do not explain the remaining spin information.

\begin{table*}
\begin{tabular}{l|c|c|c|c|c|c}
\hline\hline
    Event & \klp [bits] & \kleff [bits] & ML SNR & $m_{1,\mathrm{det}}\,[M_\odot]$ & \chieff & \chip \\
    \hline
    GW231123\_135430 & 2.28 & 3.85 & 21.2 & $172.7^{+19.0}_{-21.8}$ & $0.27^{+0.24}_{-0.35}$ & $0.76^{+0.17}_{-0.20}$ \\
    GW241011\_233834 & 1.62 & 2.69 & 36.0 & $20.4^{+2.7}_{-2.2}$ & $0.50^{+0.06}_{-0.04}$ & $0.38^{+0.13}_{-0.12}$ \\
    GW191109\_010717 & 0.97 & 1.29 & 18.2 & $79.9^{+10.2}_{-8.3}$ & $-0.31^{+0.53}_{-0.32}$ & $0.55^{+0.33}_{-0.34}$ \\
    GW190517\_055101 & 0.59 & 1.07 & 11.7 & $53.8^{+14.6}_{-10.3}$ & $0.44^{+0.21}_{-0.29}$ & $0.62^{+0.27}_{-0.34}$ \\
    GW240920\_124024 & 0.53 & 0.76 & 37.2 & $44.6^{+3.9}_{-2.8}$ & $-0.04^{+0.09}_{-0.09}$ & $0.55^{+0.29}_{-0.36}$ \\
    GW190412\_053044 & 0.37 & 1.52 & 20.1 & $34.2^{+5.7}_{-4.7}$ & $0.25^{+0.10}_{-0.10}$ & $0.20^{+0.16}_{-0.11}$ \\
    GW191204\_171526 & 0.31 & 0.40 & 17.9 & $13.8^{+3.7}_{-2.3}$ & $0.16^{+0.09}_{-0.05}$ & $0.44^{+0.37}_{-0.28}$ \\
    GW230814\_230901 & 0.27 & 1.12 & 42.3 & $36.0^{+3.4}_{-2.3}$ & $-0.01^{+0.05}_{-0.06}$ & $0.18^{+0.29}_{-0.15}$ \\
    GW231226\_101520 & 0.27 & 0.48 & 33.9 & $49.2^{+5.4}_{-3.3}$ & $-0.07^{+0.09}_{-0.09}$ & $0.38^{+0.35}_{-0.28}$ \\
    GW241225\_042553 & 0.26 & 0.33 & 17.3 & $14.1^{+3.5}_{-2.4}$ & $0.12^{+0.09}_{-0.05}$ & $0.43^{+0.45}_{-0.29}$\\

\hline\hline
\end{tabular}
\caption{Events in GWTC-5 with $\gtrsim0.25$ bits of information in component spins beyond $\chieff,\chip,$ and $q$. From left to right, the event name, information beyond $(\chieff,\chip,q)$, information beyond $(\chieff,q)$, ML matched-filter SNR, detector-frame primary mass, \chieff, and \chip. Parameters are quoted as the median and 90\% symmetric credible region. We estimate the 1-$\sigma$ uncertainty in each KL to be $\lesssim0.04$ bits (App.~\ref{app:classifier}).}
\label{tab:real-events}
\end{table*}

\subsection{Primary spin typically has the most \texorpdfstring{\spinformation}{spinformation}}

Having established that $\sim20\%$ of events have significant beyond-\chieff information in their joint spin posterior, we now turn to which component parameters tend to be better measured. For $\sim80\%$ of events, the component parameter with the most information beyond \chieff is the primary spin magnitude. This corroborates the finding in Ref.~\cite{Miller:2024sui} that population-level spin magnitude distributions are measurable beyond \chieff.

\begin{figure}
\centering
\includegraphics[width=0.7\columnwidth]{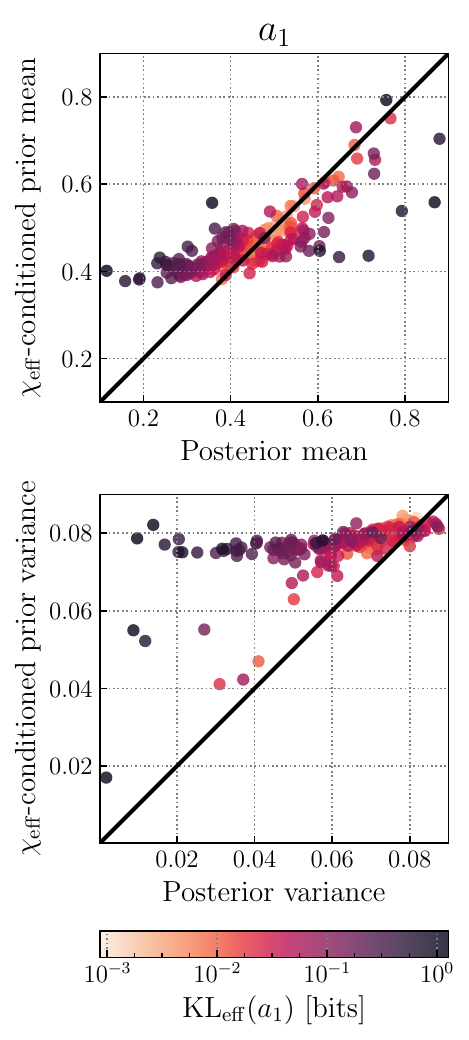}
\caption{Moments of the primary spin distribution for the full posteriors compared to the \chieff-conditioned priors, colored by beyond-\chieff information in the marginal distribution of $a_1$. The black lines mark equality.}
\label{fig:compare-moments-a1}
\end{figure}

We illustrate how beyond-\chieff information is expressed in the shape of the $a_1$ posteriors in Fig.~\ref{fig:compare-moments-a1}, which compares the first two moments of the posteriors to those of the conditioned priors. While the $a_1$ posteriors exhibit a range of means and variances, most $(\chieff,q)$-conditioned $a_1$ priors are broad and domain-covering with variances $\sim 0.08$, that of a uniform distribution on $[0,1]$. Although the conditioned prior means extend as high as $\sim 0.8$, they have a lower bound of $\sim 0.4$. This is because \chieff constrains $a_1$ asymmetrically: while a $(\chieff,q)$ pair can rule out zero spin, it cannot rule out maximal spin. At a given $(\chieff,q)$, the lower limit on primary spin is $a_1 \geq|\chieff|(1+q)-q$. If $|\chieff|\geq q/(1+q)$, then $a_1$ must be nonzero. However, $a_1$ may always be 1, since $\cos\tau_1$ can modulate their product to lie anywhere in $[-1,1]$.

Put another way, one cannot conclude a BH is nonspinning from \chieff alone since zero \chieff is degenerate with maximal in-plane spin. Lack of precession is required to conclude zero spin. For these reasons, narrow $a_1$ posteriors---particularly with low means---are not explained by \chieff and $q$, as evidenced by the correlation between events that deviate from the lines of equality in Fig.~\ref{fig:compare-moments-a1} and \kleff. We find the additional information required to claim low spin is well-captured by \chip: the median $\mathrm{KL_p}(a_1)$ is $2\times10^{-3}$ bits, and primary spin is the most informative component parameter beyond \chieff and \chip for only 10\% of events.

\subsection{Case studies of individual events}

\begin{figure*}
\centering
\includegraphics[width=0.78\textwidth]{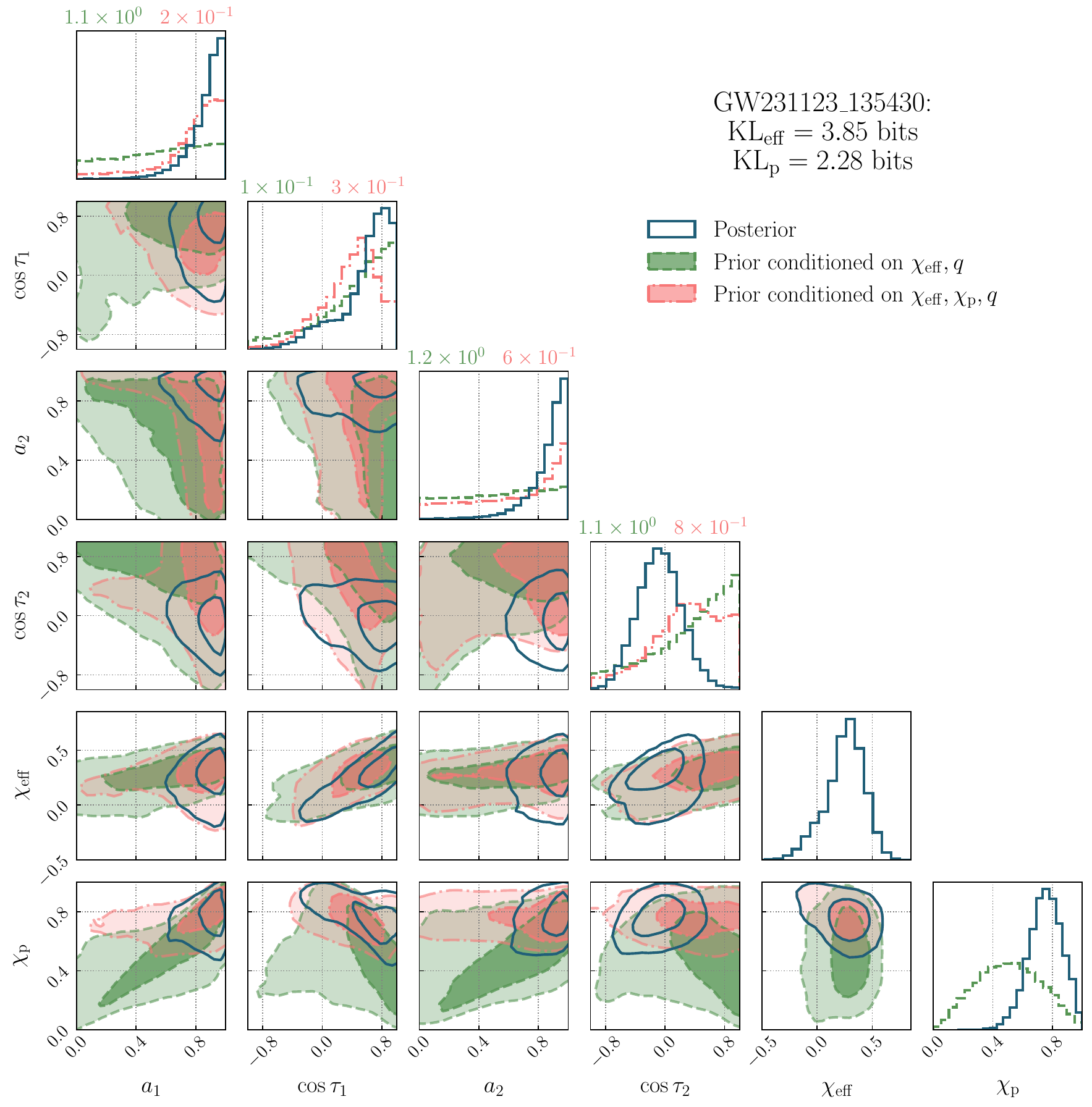}
\caption{Comparison of the posterior (blue) on spin parameters to the \chieff-conditioned prior (green) and $(\chieff,\chip)$-conditioned prior (pink) for GW231123. All distributions of \chieff agree by construction. Contours enclose 50\% and 90\% credible regions. The axis titles provide the marginal information beyond-\chieff (green) and beyond-(\chieff,\chip) (pink) in bits.}
\label{fig:spin-corner-GW231123}
\end{figure*}

We now explore three instructive events' \spinformation in detail: GW231123, GW241011, and GW250114. For each event, we describe what the \chieff and $q$ measurements imply about about the component spins, how \chip modulates those results, and whether and where the effective parameters fall short of explaining the posterior. Figures.~\ref{fig:spin-corner-GW231123}--\ref{fig:spin-corner-GW250114} visualize the constraints on component spins set by the effective parameters compared to the full posteriors. In each figure, the posterior is in unfilled blue, the prior conditioned on \chieff and $q$ is in green, and the prior also conditioned on \chip is in pink. 

\subsubsection{GW231123}

Likely the highest-mass BBH observed as of GWTC-5, GW231123 has support for extremal spin magnitudes and orbital precession~\cite{LIGOScientific:2025rsn}, which is also supported over an eccentric interpretation~\cite{Jan:2025zcm}. GW231123 claims the title of highest \spinformation in GWTC-5, cf. Table~\ref{tab:real-events}. With $\sim 3.8$ ($\sim 2.3$) bits of component spin information beyond \chieff (\chieff and \chip), it is strongly informative about component spins beyond effective parameters under a standard, unlensed quasi-circular parameter estimation hypothesis. The conclusion of beyond-effective information persists when using alternative waveforms, though the magnitude of information varies (App.~\ref{app:waveforms}).

Figure~\ref{fig:spin-corner-GW231123} shows the constraints on GW231123's component spins set by the effective parameters compared to the posterior. GW231123 has a mass ratio posterior consistent with unity, so its conditioned priors should be similar on the primary and secondary. Indeed, the green $(\chieff,q)$-conditioned marginals on $a_1$ and $\cos\tau_1$ are comparable to those on $a_2$ and $\cos\tau_2$. The  $(\chieff,\chip, q)$-conditioned tilt priors show some discrepancy since $q$ is not measured to be identically one.

GW231123 has a broad, preferentially positive \chieff distribution that is consistent with zero~\cite{LIGOScientific:2025rsn}. Together, such \chieff and $q$ measurements place minimal constraints on spin magnitudes, while the posterior instead constrains them to be rapid, and is therefore informative beyond \chieff. Due to the weight toward positive \chieff, the conditioned priors on the tilts favor alignment.

GW231123 was measured to be precessing with $\chip>0.55$ at 95\% credibility~\cite{LIGOScientific:2025rsn}. Conditioning on its high precession increases support for extremal spin magnitudes and strengthens the anticorrelation between $a_1$ and $a_2$. While a small \chieff permits both spin magnitudes to be small, a high \chip forces either or both spin magnitudes to be near-maximal. The precession constraint has nontrivial implications for the tilts, with support for exact alignment reduced in both components.

However, the data in total constrain component spins better than the effective parameters. The posterior has stronger peaks at extremal spin magnitudes than either conditioned prior. The greatest difference is in $\cos\tau_2$: while effective parameters support alignment, the data rule it out, and prefer in-plane spin. We further discuss the mass dependence of \spinformation in Sec.~\ref{ssec:origin}.

\subsubsection{GW241011}

\begin{figure*}
\centering
\includegraphics[width=0.78\textwidth]{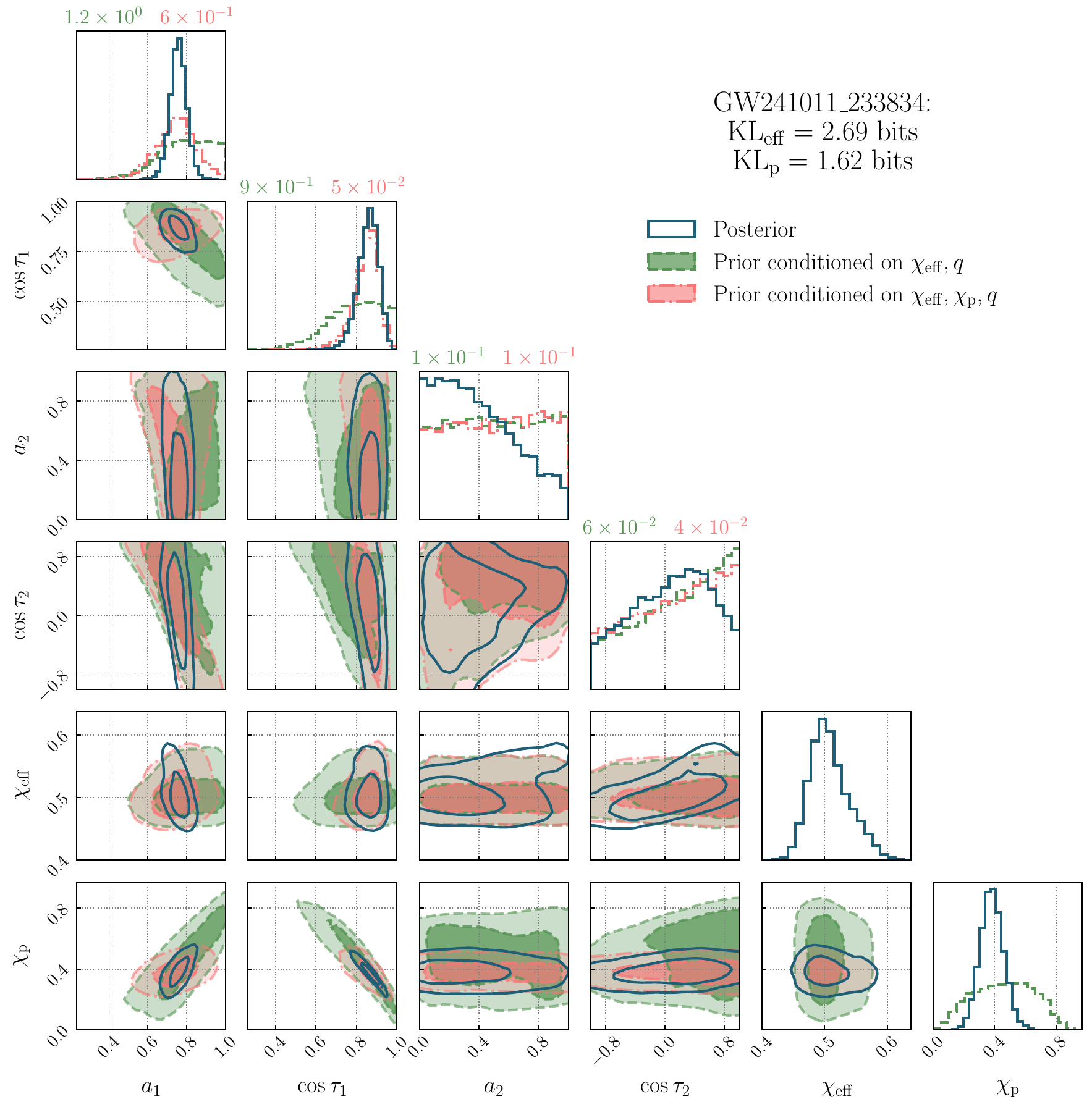}
\caption{Comparison of the posterior (blue) on spin parameters to the \chieff-conditioned prior (green) and $(\chieff,\chip)$-conditioned prior (pink) for GW241011. All distributions of \chieff agree by construction. Contours enclose 50\% and 90\% credible regions. The axis titles provide the marginal information beyond-\chieff (green) and beyond-(\chieff,\chip) (pink) in bits.}
\label{fig:spin-corner-GW241011}
\end{figure*}

GW241011 boasts the highest lower limit on primary spin magnitude of BBHs in GWTC-5~\cite{LIGOScientific:2025brd, LIGOScientific:2026sit}. With posterior comparison in Fig.~\ref{fig:spin-corner-GW241011}, GW241011 holds the second-highest \kleff and \klp in the catalog: its component spin posterior contains $\sim2.7$ bits of information beyond \chieff and $\sim1.6$ bits beyond \chieff and \chip.

GW241011 has a well-measured $q\simeq0.3$ and $\chieff\simeq 0.5$, which imply a rapid, aligned primary spin~\cite{LIGOScientific:2025brd}. In fact, every posterior sample of $(\chieff,q)$ requires $a_1\gtrsim0.2$ and $\cos\tau_1\gtrsim0.2$. Since $\chieff\simeq a_1\cos\tau_1$ at low mass ratio, $a_1$ and $\cos\tau_1$ are strongly anticorrelated in the conditioned prior. Being mass-weighted, effective parameters do not constrain the secondary's properties at low $q$. Indeed, both conditioned priors are uninformative about $a_2$. They mildly prefer positive $\cos\tau_2$, which comes from the overall positive \chieff.

Including GW241011's moderate $\chip\simeq0.4$ tightens the conditioned prior on $a_1$ and $\cos\tau_1$. Support for $a_1=1$ declines, as its measured precession is not amply extreme for maximal spin. While GW241011's \chieff allows for perfect alignment of the primary BH, its nonzero precession obliges $\cos\tau_1$ to be misaligned.

As with GW231123, the full posterior contains more \spinformation. GW241011's primary spin is measured 0.6 bits more precisely than the effective parameters require. However, its primary tilt measurement is reproducible from \chieff and \chip. Its secondary spin posteriors also contain structure beyond that imposed by effective parameters, indicating further information in the data about the secondary BH.

\subsubsection{GW250114}

The loudest GW event to date~\cite{LIGOScientific:2026wfs}, GW250114 also holds an unusual \spinformation record. While it has the third most information beyond \chieff in the catalog, almost all of it is explained by \chip: GW250114 has the largest {difference} between its beyond-\chieff and beyond-(\chieff,\chip) information.
Despite carrying $\sim2.2$ bits of spin information beyond \chieff, GW250114 has merely 0.08 bits of information beyond both effective parameters.

\begin{figure*}
\centering
\includegraphics[width=0.78\textwidth]{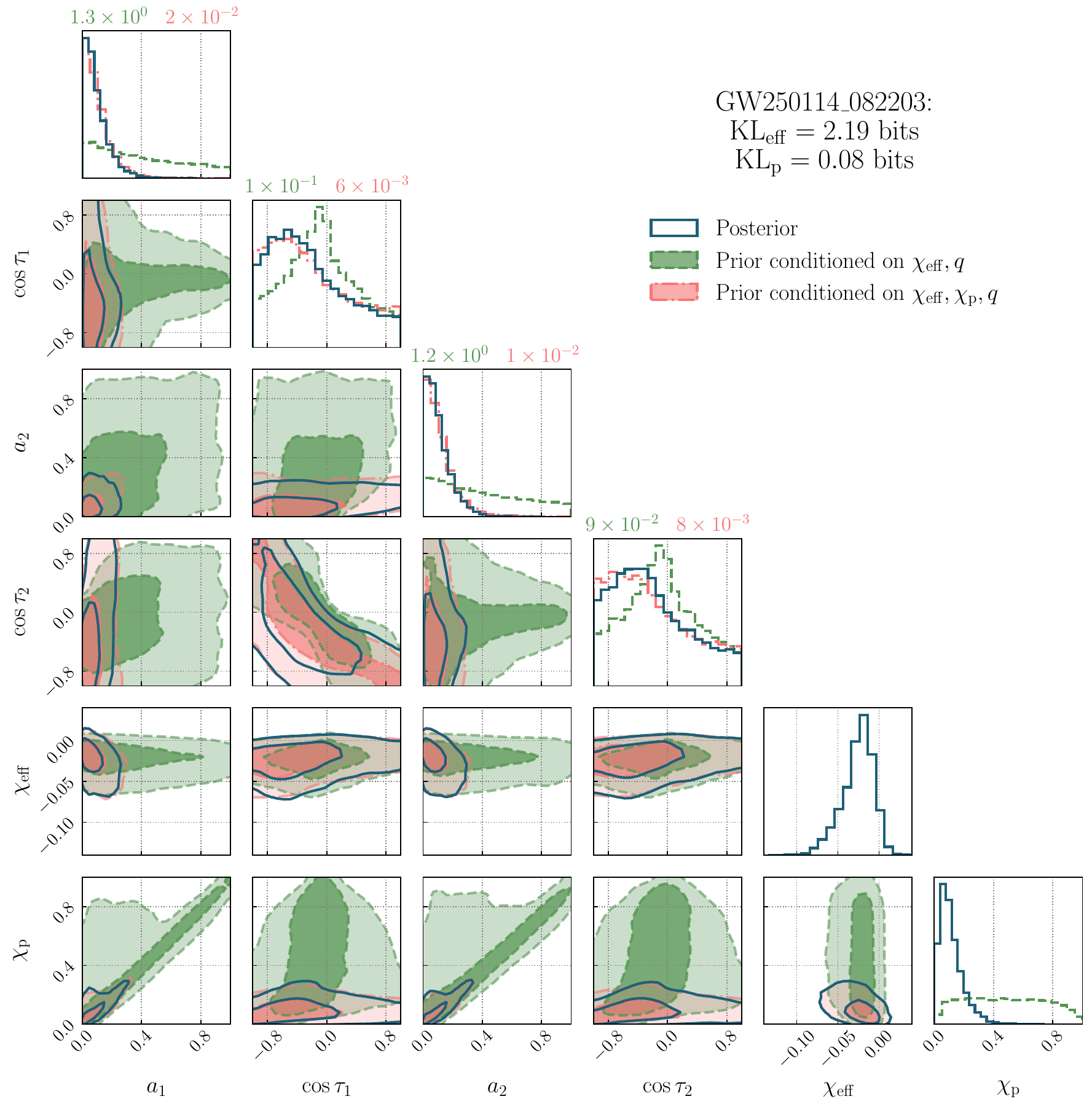}
\caption{Comparison of the posterior (blue) on spin parameters to the \chieff-conditioned prior (green) and $(\chieff,\chip)$-conditioned prior (pink) for GW250114. All distributions of \chieff agree by construction. Contours enclose 50\% and 90\% credible regions. The axis titles provide the marginal information beyond-\chieff (green) and beyond-(\chieff,\chip) (pink) in bits.}
\label{fig:spin-corner-GW250114}
\end{figure*}

Figure~\ref{fig:spin-corner-GW250114} compares its posterior to its conditioned priors. GW250114 is measured to have $\chieff\simeq 0$ and $q\lesssim1$. As in the example in Sec.~\ref{sec:cond-prior}, such a $(\chieff,q)$ pair permits high spin magnitudes so long as the associated tilt angle is nearly in-plane. The $(\chieff,q)$-conditioned prior exhibits the expected ``T'' correlation between each BH's magnitude and tilt, as well as the ``hourglass'' inversely relating the tilts. However, high spins are in contrast with the posterior, which prefers small spins. GW250114 thus contains copious information beyond \chieff.

While \chieff allows for maximal spins, GW250114's measured minimal precession ($\chip\simeq0.1$) eliminates this possibility: its spin magnitudes must be small and tilts more likely misaligned than in-plane. The constraints implied by \chieff,\chip, and $q$ are tantamount to the posterior itself. Unlike in the cases of GW231123 and GW241011, for which the effective spins were not sufficient statistics, GW250114's component spin measurements are virtually entirely described by the two effective parameters. This may be driven in part by the fact that it is slowly spinning, so its spins are more difficult to measure and its tilts have minimal meaning. The contrast between its high SNR and little \spinformation makes GW250114 an intriguing counterexample to the ncorrelation between \spinformation and SNR explored in Sec.~\ref{ssec:origin}.

\subsection{When is there \texorpdfstring{\spinformation}{spinformation}?}\label{ssec:origin}

The apparent \spinformation is of mysterious provenance. The clearest link between events with \spinformation is loudness: six of the 10 events with \klp$\gtrsim0.25$ bits have SNR $>20$, whereas 20 out of all 259 events have SNR $>20$. Beyond SNR, the informative events are otherwise dissimilar, ranging from heavy, equal-mass mergers with zero \chieff like GW230814\_230901; to asymmetric, rapidly spinning binaries like GW241011 (Tab.~\ref{tab:real-events}).

\begin{figure}
\centering
\includegraphics[width=0.7\columnwidth]{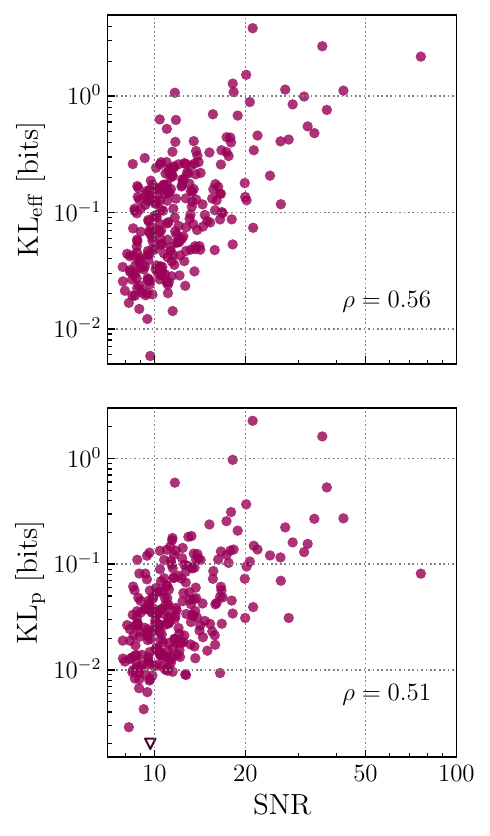}
\caption{Beyond-effective information compared to ML SNR. The top panel shows information beyond \chieff, while the bottom panel shows information beyond \chieff and \chip. The triangle represents an upper limit for GW190803\_022701, an event whose \klp estimate is $<0$.}
\label{fig:correlation-snr}
\end{figure}

Across the catalog, ML SNR correlates with both \kleff and \klp with Spearman rank-correlation coefficients of $\rho\simeq0.5$ (Fig.~\ref{fig:correlation-snr}). An overall relation between SNR and information is anticipated: effective parameters are constructed from leading-order inspiral terms, whereas beyond-effective information must lie in subdominant effects, which are better-resolved in louder signals~\cite{Wolfe:2025coy}.
The relation between SNR and \kleff is roughly log-linear with significant noise; we find information scales as $\mathrm{KL_{eff}}\propto\mathrm{SNR}^{2.1\pm0.2}$.
In the top panel of Fig.~\ref{fig:correlation-snr}, the rising lower envelope and vacant lower right corner indicates that, among observed events, there is an information floor at a given SNR. The lowest \kleff in an SNR $>30$ event is $\sim0.5$ bits. The upper envelope, in contrast, remains hazier. SNR is a strong predictor of a lower limit in beyond-\chieff information, but does not clearly imply an information maximum.

\klp also exhibits a noisy relationship with SNR, scaling as $\mathrm{KL_{p}}\propto\mathrm{SNR}^{1.7\pm0.2}$. Its lower envelope never reaches our threshold of significantly informative; only four of the seven events with SNR $>30$ have \klp$>0.25$ bits. GW250114 most strongly breaks this trend. At SNR $\sim 75$ and \klp$=0.08$, GW250114 lies well below the envelope of the remainder of the data.

Just as the informative events vary in their likely properties, so do the three weakly informative SNR $>30$ events span parameter space. The aforementioned GW250114 has equal, moderate masses and low spins; GW240925\_005809 is low-mass with near-zero \chieff; and GW241127\_061008 is heavy, asymmetric, and precessing. Loudness is not sufficient to produce \spinformation beyond both effective parameters, nor do other binary properties explain these events' lack of information. Moreover, loudness is not required: GW190517\_055101 has SNR $\simeq12$ and $\sim0.6$ bits of posterior information beyond \chieff and \chip. 

Alongside SNR, ML detector-frame primary mass is the strongest predictor of \spinformation (both $\rho\simeq-0.5$).\footnote{ML redshift correlates as strongly, but is not an independent driver: distant sources are preferentially quiet and high-mass, both of which tend to suppress \spinformation.} On one hand, this anticorrelation may be attributable to the accumulation of inspiral and precession cycles in-band at lower masses. More cycles close to merger can improve measurements of effective parameters~\cite{Ng:2018neg, Green:2020ptm, Biscoveanu:2026ikx}
and likely also of the constituent spin components (in the extremely early inspiral, the precession timescale is so long that precession is inconsequential in the waveform~\cite{Mangiagli:2018kpu}). 

On the other hand, since \chieff is a leading-order spin term in the inspiral, for massive events dominated by merger and ringdown, \chieff (and \chip) are not accurate representations of the spin dynamics in the waveform and should not be expected to summarize spin constraints~\cite{Biscoveanu:2021nvg, Miller:2023ncs, Miller:2025eak, Kang:2025nio}. GW231123 supports this second interpretation: likely the highest-mass observed binary, it the most \spinformation despite essentially zero in-band inspiral. All its spin information must therefore lie in other components of the signal. The case of GW231123 implies component spins can be effectively extracted from merger and ringdown~\cite{Biscoveanu:2021nvg, Miller:2025eak}; however, the signal's extremal nature on the edge of waveform-calibration validity make it unwise to generalize its \spinformation properties to the entire population. These contrasting observations imply what drives beyond-effective spin measurement is not reducible to a simple marginal surface or scaling argument.

Combined, the correlations between \spinformation, SNR, and mass imply well-resolved inspiral cycles (as seen in loud, low-mass events) contribute to \spinformation, but are neither necessary nor sufficient. None of the established drivers of spin measurement---neither spin magnitude, mass asymmetry, binary orientation, nor combinations thereof---explain the observed trends. While the real catalog precludes a systematic analysis of how \spinformation scales with intrinsic parameters, the lack of correlations may also suggest that what controls precise spin measurement does not necessarily correspond to what drives beyond-(\chieff,\chip) information. No low-dimensional cut in posterior parameter space cleanly separates events with \spinformation from uninformative ones.  

\subsection{\texorpdfstring{\spinformation}{Spinformation} can be noise-driven}\label{ssec:noise}

\spinformation is a property of a posterior, not directly of true source parameters, and thus depends on the combined effects of signal and noise. Beyond-effective structure may therefore be influenced by the particular noise realization. To assess this dependence, we take a simulated catalog of 1600 BBHs~\cite{Wolfe:2026wgk} and re-analyze 10 events possessing high \spinformation (\klp) with six different realizations of Gaussian noise. Selected source properties of the 10 simulated signals and their initial KLs are given in Tab.~\ref{tab:injections}.

\begin{table}[]
\begin{tabular}{l|c|c|c|c|c|c|c}
\hline\hline
Event & SNR & $m_1\,[M_\odot]$ & $q$ & $\chi_\mathrm{eff}$ & $\chi_\mathrm{p}$ & \begin{tabular}[c]{@{}c@{}}\kleff\\ {[}bits{]}\end{tabular} & \begin{tabular}[c]{@{}c@{}}\klp\\ {[}bits{]}\end{tabular} \\ \hline
43   & 102.7 & 32.5 & 0.78 & $-0.06$ & 0.25 & 3.35 & 0.87 \\
184  & 17.4 & 32.1 & 0.38 & 0.15    & 0.89 & 1.65 & 0.22 \\
203  & 21.9 & 35.6 & 0.93 & 0.37    & 0.49 & 0.36 & 0.25 \\
265  & 21.3 & 31.8 & 0.55 & $-0.11$ & 0.67 & 1.93 & 0.30 \\
364  & 39.3 & 39.5 & 0.54 & 0.30    & 0.22 & 1.10 & 0.55 \\
443  & 24.1 & 41.4 & 0.76 & 0.18    & 0.35 & 0.32 & 0.25 \\
664  & 38.8 & 28.0 & 0.54 & 0.27    & 0.30 & 1.04 & 0.53 \\
1532 & 27.5 & 36.2 & 0.40 & $-0.04$ & 0.09 & 0.76 & 0.24 \\
1554 & 13.0 & 34.4 & 0.41 & 0.34    & 0.33 & 1.00 & 0.51 \\
1599 & 84.2 & 51.1 & 0.81 & 0.30    & 0.22 & 0.88 & 0.58 \\ \hline\hline
\end{tabular}
\caption{Selected properties of simulated events used to assess the impact of noise on \spinformation. The event indices match Ref.~\cite{Wolfe2025_zenodo17080422}. The KLs are from the original noise realization.}
\label{tab:injections}
\end{table}

Over the 10 injections, the spread in \kleff across noise realizations ranges from 0.4 to $>1.5$ bits. The scatter in \klp is as low as 0.1 and up to 0.65 bits. The noise spreads in \kleff and \klp are not correlated. Since we tested a limited number of noise realizations on few events, selected for their \spinformation, these spreads due to noise are not authoritative for all signals but rather representative of what noise can induce.

\begin{figure}
\centering
\includegraphics[width=0.8\columnwidth]{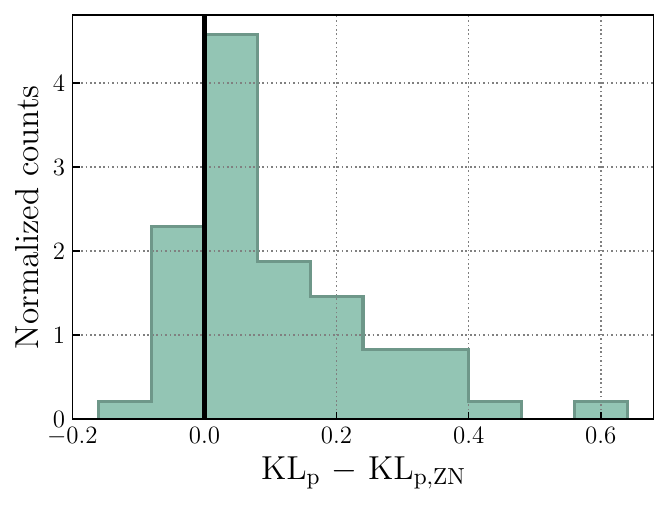}
\caption{From 10 simulated signals, the distribution of the difference between the beyond-effective information \klp from noisy data realizations and that from a zero-noise realization of the same signal. The distribution stacks six noise realizations for each of the 10 different signals.}
\label{fig:compare-kl-noise}
\end{figure}

One method to characterize the ``intrinsic'' \spinformation in a signal is via the posterior from a zero-noise data realization. Analyzing the 10 signals in zero noise, we find that the zero-noise (\chieff,\chip)-spinformation is generally lower than that from noisy data realizations. We illustrate this in Fig.~\ref{fig:compare-kl-noise}, which shows the distribution of the difference between the beyond-effective information \klp from noisy data realizations and the zero-noise realization of each signal. The positive mean and long positive tail indicate that these noisy data realizations on average produce posteriors with more beyond-effective information.

Finally, we observe that the informative posteriors have smaller variances than the conditioned priors, also seen in the real events. The posteriors are not usually narrowed around the truth, however. Instead, their modes are often simultaneously biased, as is typical in noisy data, such that the truths lie at comparable percentiles in the posteriors and conditioned priors.

We identify two takeaways regarding the effect of noise: first, information content \kleff$<1$ bit and \klp$<0.5$ bits may be driven more by noise than signal properties. Second, beyond-effective information does not necessarily imply a more accurate measurement of true parameters.

\section{Conclusions}\label{sec:conclusion}

We assessed whether the effective parameter reduction (\chieff, \chip) can summarize current component spin constraints and found:

\begin{enumerate}

    \item \textbf{Primary spins are often measured better than $\boldsymbol{\chieff}$ alone can explain.} Effective inspiral spin alone indeed throws out information regarding spin magnitudes, but usually well-describes alignment constraints.
    
    \item \textbf{The two effective spins almost completely describe the catalog's component spin measurements.} Using a noise-informed threshold, we argue only five events have component spin constraints not captured by \chieff and \chip that, further, are confidently not driven by noise.

    \item \textbf{Noise can fluctuate \spinformation by over a bit}. \spinformation is not a property of the signal, but a metric computed from an inference result on data. Simulations suggest it is frequently dominated by noise.

\end{enumerate}

Where \spinformation exists, its origin is difficult to identify. We defer a systematic analysis of how \spinformation scales with intrinsic parameters and higher-order post-Newtonian terms to future work.

While in this study we focus on prior conditioning on \chip, other works have proposed additional ways to quantify and describe precession that offer complementary approaches to analyzing spin measurement. For instance, the precessing SNR~\cite{Fairhurst:2019vut, Fairhurst:2019srr, Green:2020ptm} may be able to predict the measurability of precession. Ref.~\cite{Gerosa:2015tea} argues precession can be encoded into four discrete morphologies, the observed distribution of which has been probed in Refs.~\cite{Gangardt:2022ltd, Johnson-McDaniel:2023oea}. Employing a range of methods to study spin constraints will enable thorough comprehension of both the underlying physics and how the data encode it.

In the \spinformation framework, the minimal information beyond effective parameters in the catalog implies individual-event component spin posteriors are almost always derivable from effective spin ones. Their geometry is enforced by \chieff and \chip, coupled with mass ratio. 

Population-level inference can aggregate poorly-measured events into a better constraint on the population. However, companion work \cite{Wolfe:2026pjn} suggests that component spin population results indeed extrapolate constraints from effective spins. They find that the lack of beyond-effective information in the population has interpretative implications: for example, tentative evidence for a preferred BH spin orientation can be traced to population modeling assumptions as opposed to robust GW spin measurements. In turn, any features identified in component-spin models that do not propagate to distinct joint-effective-space distributions are likely model-, prior- or Poisson-noise-driven.

Looking ahead, component spin measurements for individual events should be viewed equivalently to joint effective constraints---and caution must be taken before making population claims that effective parameters cannot reproduce.

\section{Acknowledgments}

We thank Sofía Álvarez-López, Sylvia Biscoveanu, Mark Hannam, and Aditya Vijaykumar for discussions.
C.P. and N.E.W. are supported by the National Science Foundation Graduate Research Fellowship under grant no. DGE-2141064.
S.V. is partially supported by the NSF grant no. PHY-2045740.
D.G. is supported by
ERC Starting Grant No.~945155--GWmining, 
Cariplo Foundation Grant No.~2021-0555, 
Italian-French University (UIF/UFI) Grant No.~2025-C3-386,
MUR Grant ``Progetto Dipartimenti di Eccellenza 2023-2027'' (BiCoQ),
MUR Young Researchers Grant No. SOE2024-0000125. 
and the INFN TEONGRAV initiative.
The Flatiron Institute is a division of the Simons Foundation.
The authors are grateful for computational resources provided by the LIGO Laboratory supported by National Science Foundation grants no. PHY-0757058 and no. PHY-0823459.
This material is based upon work supported by NSF's LIGO Laboratory which is a major facility fully funded by the National Science Foundation and has made use of data or software obtained from the Gravitational Wave Open Science Center (gwosc.org), a service of the LIGO Scientific Collaboration, the Virgo Collaboration, and KAGRA.

\bibliography{paper}

\appendix

\section{Deriving the conditional priors}\label{app:cond-prior-chieff}

Conditioning an isotropic component prior on an effective spin and mass ratio induces a three-dimensional constraint manifold embedded in the four-dimensional component spin space. We obtain a level set of the map $f:[0,1]^2\times[-1,1]^2\to[-1,1]: f(a_1,\cos\tau_1,a_2,\cos\tau_2\mid q) = (a_1\cos\tau_1+qa_2\cos\tau_2)/(1+q)$; i.e., the preimage $f^{-1}(\chieff\mid q)$. 
We derive the $(\chieff,q)$-conditioned prior [Eq.~(\ref{eq:cond-prior})] in two stages. First, we change variables to the orbit-aligned components of the spins, $\cz i = a_i\cos\tau_i,\ i\in[1,2]$, and compute the induced prior. Second, we derive the density on spin magnitude and direction conditioned on the aligned spin.

In terms of aligned spin, we use Bayes' theorem to write the conditional prior on $\cz2$, from which $\cz1$ follows deterministically:
\begin{equation}\label{eq:cond-cz2}
    \begin{aligned}
        p(\cz2\mid\chieff,q,\mathrm{iso}) &\propto p(\chieff\mid\cz2,q,\mathrm{iso}) \ p(\cz2\mid\mathrm{iso})\\
        & \propto p(\cz1 = \chieff(1 + q) - q\cz2\mid\mathrm{iso})\\
        &\times p(\cz2\mid\mathrm{iso}).
    \end{aligned}
\end{equation}
For equality, the first line would include a normalization factor $p(\chieff\mid q,\mathrm{iso})$, and the second line the Jacobian $|\dd \cz1 / \dd \chieff| = 1 + q$. As these are independent of the component spins, they are absorbed into the normalization. In addition, conditioning $\cz2$ on \chieff and $q$ sets its possible range: an extremal $\cz1$ limits $\cz2\in\left[\frac{\chieff(1+q)-1}{q},\,\frac{\chieff(1+q)+1}{q}\right]\cap[-1,1]$. 

Each marginal prior $p(\cz i\mid\mathrm{iso})$ is obtained by marginalizing $p(\cz i\mid a_i,\mathrm{iso})$ over $a_i$. The change of variables $(a_i, \cos\tau_i)\to (a_i, \cz i)$ yields
\begin{equation}\label{eq:joint-ai-czi}
    p(a_i,\cz i\mid\mathrm{iso}) = p(a_i,\cos\tau_i\mid\mathrm{iso})\,\left|\frac{\dd\cos\tau_i}{\dd\cz i}\right|=\frac{1}{2a_i}
\end{equation} and the conditional $p(\cz i\mid a_i,\mathrm{iso})$ follows from Bayes' theorem. The boundary $\cos\tau_i\in[-1,1]$ requires $a_i\in[|\cz i|, 1]$, so
\begin{equation}
\begin{aligned}\label{eq:marginal-pc}
    p(\cz i\mid\mathrm{iso}) &= \int_{|\cz i|}^1\dd a_i\, p(\cz i\mid a_i, \mathrm{iso})\,p(a_i\mid\mathrm{iso})\\
    &= \int_{|\cz i|}^1 \dd a_i\,\frac{1}{2 a_i}= -\frac12\ln|\cz i|.
\end{aligned}
\end{equation}
Substituting Eq.~(\ref{eq:marginal-pc}) into Eq.~(\ref{eq:cond-cz2}), we obtain
\begin{equation}
\begin{aligned}\label{eq:draw-aligned}
    p(\cz2\mid\chieff,q,\mathrm{iso}) &\propto \ln|\chieff(1+q)-q\cz2|\\
    &\times\ln|\cz 2|,
\end{aligned}
\end{equation}
with $\cz2$ in the range given above and normalized accordingly.

We now handle the hyperbolic constraint on spin magnitudes and directions. The conditional $p(a_i \mid \cz i,\,\mathrm{iso})$ follows from Eq.~\ref{eq:joint-ai-czi} and is
\begin{equation}\label{eq:draw-mags}
    p(a_i\mid \cz i,\,\mathrm{iso}) = \frac{1}{-a_i \ln|\cz i|},\quad a_i\in[|\cz i|, 1].
\end{equation} After fixing $a_i$, the tilt $\cos\tau_i$ is set by $\cos\tau_i = \cz i / a_i$. In practice, we sample from Eqs.~\ref{eq:draw-aligned} and~\ref{eq:draw-mags} using inverse CDF methods.

\begin{figure*}
\centering
\includegraphics[width=0.7\textwidth]{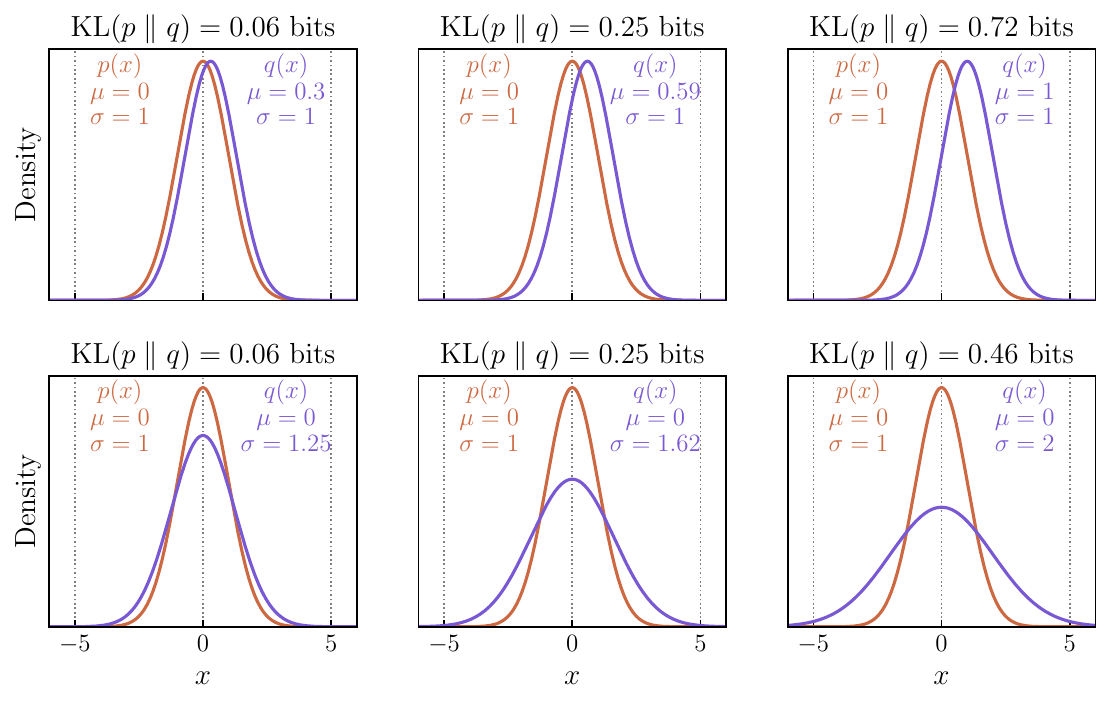}
\caption{Examples of the KL divergence between univariate Gaussians offset in mean (top row) and scale (bottom row).}
\label{fig:illustrate-kl}
\end{figure*}

\subsection{Incorporating the precessing spin constraint}\label{app:cond-prior-chip}

We now extend to the prior on component spins conditioned on both effective spin parameters [Eq.~(\ref{eq:cond-chip-prior})]. In short, we reuse the aligned-component prior [Eq.~(\ref{eq:draw-aligned})], draws from which satisfy \chieff by construction, and employ importance sampling to select pairs by their compatibility with \chip. We first compute the induced prior on the aligned spin components (conditioned on both effective parameters), followed by the in-plane components. We drop the explicit initial isotropic prior on the right-side of the conditional bars.

Analogous to the aligned components $\cz i=a_i\cos\tau_i$, define the in-plane spin components
\begin{align}
    \cperp i = a_i\sin\tau_i\in[0,1].
\end{align}
Note that $a_i^2 = \cz i^2 + \cperp i^2$. In this coordinate system, \chieff only explicitly depends on $(\cz1,\cz2)$, and \chip on $(\cperp1,\cperp2)$. We write the conditional prior as

\begin{equation}\label{eq:joint-prior-chip}
\begin{aligned}
  &p(\cz1,\cperp1,\cz2,\cperp2\mid\chieff,\chip,q)\\
&=p(\cz2 \mid\chieff,\chip, q)\,p(\cperp1,\cperp2\mid\cz1,\cz2,\chip,q).
\end{aligned}
\end{equation} One of the in-plane components in the second term is fixed by \chip; which component depends on which BH dominates \chip, with probabilities computed below.
The aligned term further factorizes via Bayes' theorem:
\begin{equation}
\begin{aligned}
    &p(\cz2\mid\chieff,\chip, q) \\
    & \propto p(\cz2\mid\chieff,q) \, p(\chip\mid\cz2,\chieff, q).
\end{aligned}
\end{equation}
The first term is Eq.~(\ref{eq:draw-aligned}); the second term is the weight $w(\cz1,\cz2)$, the prior density of \chip given the aligned spins (we note that the above Bayes' expansion is in terms of \chieff, but can be written equivalently in terms of $\cz1$). The weight takes the densities of the in-plane components, conditional on their aligned components, and marginalizes over the in-plane components restricted to the surface respecting \chip: 
\begin{equation}\label{eq:weight-chip}
\begin{aligned}
    w(\cz1,\cz2) &= \int p(\cperp 1\mid \cz 1)\, p(\cperp 2\mid \cz 2)\\
    &\times \delta[\max(\cperp1,\tilde q\cperp2)-\chip]\,\dd\cperp1\dd\cperp2,
\end{aligned}
\end{equation} where $\tilde q = q (3+4q)/(4+3q)$. Each prior $p(\cperp i\mid\cz i)$ is
\begin{align}\label{eq:p-perp-given-aligned}
    p(\cperp i\mid\cz i) = \frac{1}{-\ln|\cz i|}\frac{\cperp i}{\cz i^2 + \cperp i^2},
\end{align} with $\cperp i\in\left[0,\sqrt{1-\cz i^2}\right]$, which is equivalent to Eq.~(\ref{eq:draw-mags}) after the change of variables. This has CDF
\begin{align}
    F_i(x) = \frac{1}{-2\ln|\cz i|}\ln\left(1 + \frac{x^2}{\cz i^2}\right).
\end{align} The ``max'' in \chip splits Eq.~(\ref{eq:weight-chip}) into two branches: one in which the primary's spin dominates ($\cperp1 = \chip$) and one in which the secondary's does ($\cperp2 = \chip/\tilde q$). The weight in each branch is the density for that in-plane component to produce \chip---it absorbs the delta function---times the CDF of the non-dominant component up to its cap (above which it would dominate), as well as the Jacobian for $w_2$:
\begin{equation}
\begin{aligned}
    w &= w_1 + w_2,\\
    w_1 &= p(\cperp1=\chip\mid\cz1) F_2\left[\min\left(\chip/\tilde q, \sqrt{1-\cz2^2}\right)\right],\\
    w_2 &= \frac{1}{\tilde q} p(\cperp2=\chip/\tilde q\mid\cz2) F_1\left[\min\left(\chip, \sqrt{1-\cz1^2}\right)\right];
\end{aligned}
\end{equation} with $w_1$ valid if \chip is less than the maximum $\cperp1$, $\chip<\sqrt{1-\cz1^2}$ and 0 otherwise; similarly, $w_2$ holds if $\chip/\tilde q<\sqrt{1-\cz2^2}$ and is 0 otherwise.

This decomposition suggests an importance-sampling scheme: first, draw a pool of aligned components respecting \chieff via Eq.~(\ref{eq:draw-aligned}). Second, importance-sample one draw from these pairs according to probability $\propto w$. Third, sample the in-plane components (and therefore magnitudes and directions) via the second term of Eq.~(\ref{eq:joint-prior-chip}).
The sampling procedure is as follows:
\begin{enumerate}
    \item Draw $M$ candidate aligned pairs $(\cz1,\cz2)_j\sim p(\cz2\mid\chieff, q)$ from Eq.~(\ref{eq:draw-aligned}).
    \item Select one candidate with probability $\propto w_j$. Its primary spin dominates with probability $w_{1,j}/w_j$ (else secondary). The corresponding in-plane component is thus fixed; draw the other using Eq.~(\ref{eq:p-perp-given-aligned}), truncated appropriately.
    \item With $\cz i$ and $\cperp i$ in hand, map back to $(a_i,\cos\tau_i)$.
    
\end{enumerate}

\section{Illustrating information content}\label{app:info}

The KL divergence between two $d$-dimensional Gaussian distributions $p_1, p_2$ with means $\mu_1, \mu_2$ and covariance matrices $\Sigma_1,\Sigma_2$ is, in bits,

\begin{equation}\label{eq:kl-gaussians}
\begin{aligned}
    \mathrm{KL}(p_1\parallel p_2) &= \frac{1}{2\ln 2}\bigg[ \ln\frac{|\Sigma_2|}{|\Sigma_1|}+\mathrm{Tr}\left(\Sigma_2^{-1}\Sigma_1\right)\\
    &+(\mu_2-\mu_1)^T\Sigma_2^{-1}(\mu_2-\mu_1)-d\bigg].
\end{aligned}
\end{equation}
The four terms may be thought of follows: the first two account for the changes in spread and covariance between the distributions, the third penalizes offsets in mean, and the final term regularizes the KL to be zero for identical distributions. In the absence of correlations, information is additive across dimensions.

We visualize the information from univariate mean (top) and scale (bottom) shifts in Fig.~\ref{fig:illustrate-kl}. The left column shows mean and scale shifts producing $\sim0.06$ bits of information, approximately that required in each component parameter to produce a total of 0.25 bits in four dimensions, our heuristic threshold. The central column demonstrates that 0.25 bits in a single dimension is equivalent to a mean offset of $\sim $ 0.6 standard deviations, or co-located distributions with the reference $\sim 60\%$ wider.

\section{KL estimation techniques}\label{app:classifier}

The KL divergence may be written as the expected value of the log density ratio:
\begin{equation}
    \mathrm{D_{KL}}(p\parallel q) = \mathbb E_{x\sim p(x)}\left[\log \frac{p(x)}{q(x)}\right].
\end{equation} Estimating a KL divergence therefore reduces to estimating the density ratio $p(x)/q(x)$ at points $x$ on the support of $p$. Numerous methods exist for density ratio estimation; see Ref.~\cite{Sugiyama_Suzuki_Kanamori_2012} for a comprehensive treatment.

The binary classifier approach makes use of the observation that given samples from $p(x)$ labeled 1, and those from $q(x)$ labeled 0, the optimal classifier $h$ outputs $h(x) = p(x) / (p(x) + q(x))$, which can be remapped to the ratio $p(x)/q(x) = h(x) / (1 - h(x))$. We train a multi-layer perceptron to distinguish posterior samples from conditioned prior samples using \texttt{equinox}~\cite{kidger2021equinox}, accelerated by \texttt{JAX}~\cite{jax2018github}, using the \texttt{Adamw} optimizer~\cite{Kingma:2014vow, loshchilov2019, deepmind2020jax} with a cosine-decay learning rate~\cite{smith2018} to minimize the binary cross-entropy loss. We use multivariate Gaussians of similar sample size to the posteriors ($n=10^4$) as known test cases to calibrate the classifier architecture and training parameters. Longer training does not improve convergence but instead promotes memorization of the training set, which biases KL estimates upward. Similarly, highly expressive neural networks overfit. Across a range of test cases, we find a neural network width of 32 and depth of 2, trained for 5000 steps, obtains convergence without overfitting. We further ameliorate training-set memorization with five-fold cross-fitting; i.e., we train each dataset in five portions and score each block using a model trained on the remaining data. The fractional uncertainty in the KL estimate from different data realizations is order of 2\%.

On each event, we repeat the classifer KL on 20 realizations from the conditioned priors to obtain mean \kleff and \klp estimates and a coarse evaluation of their standard deviations across sample sets. The data-realization uncertainty generally scales with the information and reaches up to $\sim0.04$ bits for \kleff and $\sim0.02$ bits for \klp. We cross-check the resulting classifier KL divergence means using two additional methods: KLs from $k$-nearest neighbor~\cite[$k$-NN,][]{PerezCruzKL, WangKL} density ratio estimators and direct density fits via normalizing flow emulators~\cite{JMLR:v22:19-1028, 9089305}. For the former, we use \texttt{KDTree} in \texttt{scipy}~\cite{2020SciPy-NMeth} and test $k$ from 1 to 16. For the latter, we use block neural autoregressive flows~\cite{decao2019blockneuralautoregressiveflow} as implemented in \texttt{flowjax}~\cite{ward2023flowjax} with an autodifferential inverse~\cite{Wolfe:2026dcq} using a single layer and neural network width of 8. We again calibrate these methods with Gaussian test cases. For these consistency checks, we use a single realization of the conditioned priors; that is, we do not separately estimate the uncertainty from each method. As additional validation, we repeat the classifier estimates using a neural network width of 64 and 3000 training steps. 

The three methods produce comparable KLs for the majority of events. The distribution of the classifier minus flow \klp has mean $-2\times10^{-2}$ bits and standard deviation $2\times10^{-2}$ bits; that of the classifier minus 4-NN \klp has mean $4\times10^{-3}$ bits and standard deviation $2\times10^{-2}$ bits. The largest discrepancies are between the flows and classifier for two high-information events: GW191109\_010717, for which the classifier \kleff and \klp are $\sim0.2$ bits higher, and GW250114, for which the classifier \kleff is 0.3 bits lower. That these differences appear in high-information events is unsurprising; error in KL estimation tends to scale with the KL itself, since the estimate becomes dominated by regions where one dataset is sparse. The observed differences are fractionally small ($\sim10\%$) and do not change the conclusion that these events contain significant information.

The normalizing flows produce a higher information floor ($\sim2\times10^{-2}$ bits) than either density ratio method. For $k\lesssim4$, the $k$-NN technique returns more negative KLs than the classifier (before averaging over data realizations), a consequence of noisy mean estimation near zero. We use the classifier in the main text and note that no qualitative results---and only minor quantiative results---change if we adopt either alternative.

For the one-dimensional marginal KLs, we directly estimate densities with the bounded kernel density estimate implementation in \texttt{pesummary}~\cite{Hoy:2020vys}, which employs a reflective boundary and bandwidth determined by Scott's rule~\cite{Scott1992} in \texttt{scipy}~\cite{Virtanen:2019joe}.

\section{Posterior samples}\label{app:samples}

We use publicly-released posterior samples for the 259 BBHs with FAR $<1\,\mathrm{yr}^{-1}$ in GWTC-5. We select the \textsc{NRSur7dq4}~\cite{Varma:2019csw} waveform when available (105 events), followed by \textsc{IMRPhenomXPHM-SpinTaylor}~\cite{Colleoni:2024knd} (85 events), and finally \textsc{IMRPhenomXPHM}~\cite{Pratten:2020ceb} (69 events; all those through the Third Gravitational-Wave Transient Catalog). To minimize sampling noise in the KL computation, we use up to 25,000 posterior draws per event, randomly subsampling the posterior for events with larger numbers of samples. 

\section{GW190814 and neutron-star-containing events}\label{app:nsbh}

GW190814 was the merger of a $\sim23 M_\odot$ BH with a $\sim2.6M_\odot$ compact object of uncertain source class~\cite{LIGOScientific:2020zkf} and is usually excluded from BBH-only analyses due to the ambiguous nature of the secondary. The upper bound on its primary spin is 0.08. With $\chieff\simeq0$, GW190814 has $\sim2$ bits of information beyond \chieff, akin to GW250114 in that $\chieff=0$ does not well-constrain $a_1$. Once its near-zero \chip constraint is included, the posterior contains 0.2 bits of information beyond the (\chieff,\chip)-conditioned prior. Neither effective parameters nor posterior provide information about the secondary BH's spins, which is expected given the extreme mass ratio.

As no waveform contains both precession and tidal effects, we study neutron star--BH events using variants of \textsc{IMRPhenomXPHM}. GW200105 and GW230518 are similar to GW190814: with well-constrained low mass ratios and zero spin, \chieff alone cannot reproduce the posteriors, but the pair (\chieff,\chip) is nearly sufficient. No neutron-star-containing events have \klp$>0.25$ bits.

\section{Impact of waveform systematics}\label{app:waveforms}

\begin{figure}
\centering
\includegraphics[width=\columnwidth]{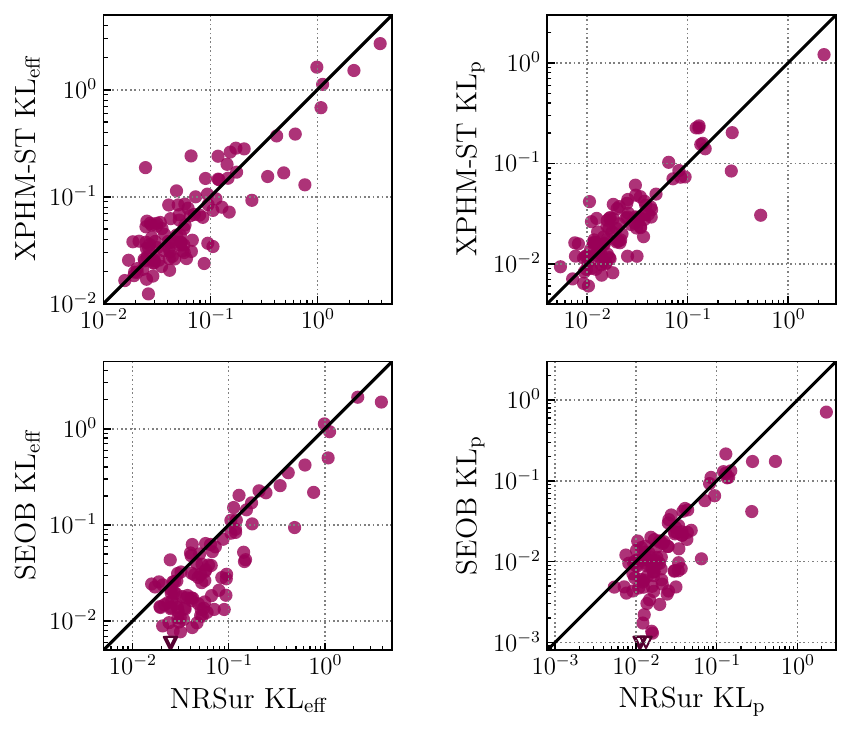}
\caption{Comparison of \kleff (left column) and \klp (right column) for catalog events analyzed with the \textsc{NRSur7dq4} waveform versus \textsc{IMRPhenomXPHM-SpinTaylor} (top row) and \textsc{SEOBNRv5PHM} (bottom row). Black diagonal lines mark equality. Waveform names are abbreviated in the figure and all values are in bits. Triangles mark upper limits for events whose KL estimates are negative.}
\label{fig:waveform-comparison}
\end{figure}

For events of sufficiently high mass to be analyzed with \textsc{NRSur7dq4}, we compare the beyond-effective information when analyzed with the surrogate waveform to that when using the alternative waveforms of \textsc{IMRPhenomXPHM-SpinTaylor} and \textsc{SEOBNRv5PHM}~\cite{Ramos-Buades:2023ehm}, shown in Fig.~\ref{fig:waveform-comparison}. Across the 105 qualifying events, the median difference in \klp between \textsc{NRSur7dq4} and \textsc{IMRPhenomXPHM-SpinTaylor} is $\sim10^{-4}$, with standard deviation 0.12.

Three events---GW231123, GW240920\_124024, and GW250114---have over half a bit more beyond-\chieff information in their \textsc{NRSur7dq4} posteriors than their \textsc{IMRPhenomXPHM-SpinTaylor} posteriors. GW231123, GW240920\_124024, and GW241225\_082815 satisfy the same criterion for \textsc{SEOBNRv5PHM}. For beyond-(\chieff,\chip) information, GW231123 has over a bit more information when analyzed with \textsc{NRSur7dq4} than either alternative; GW240920\_124024 has close to half a bit more with \textsc{NRSur7dq4} than either alternative.

For three events the surrogate waveform produced posteriors with \klp above our nominal threshold of 0.25 bits, but neither alternative waveform did: GW230814\_230901, GW231226\_101520, and GW240920\_124024. No events satisfy the alternative (i.e., a posterior with significant \spinformation under \textsc{IMRPhenomXPHM-SpinTaylor} or \textsc{SEOBNRv5PHM}, but not under \textsc{NRSur7dq4}). In a handle of cases, the higher-fidelity waveform enables considerable further recovery of component spin parameters.

\textsc{SEOBNRv5PHM} posteriors have systematically lower \spinformation than those from \textsc{NRSur7dq4} or \textsc{IMRPhenomXPHM-SpinTaylor}. Roughly 84\% (80\%) of the events have higher \kleff (\klp) using \textsc{NRSur7dq4} than \textsc{SEOBNRv5PHM}. In contrast, both figures are $\sim50\%$ for \textsc{IMRPhenomXPHM-SpinTaylor}.
Whether the information decline with \textsc{SEOBNRv5PHM} is significant, and if so, precisely what content in the surrogate waveform lacked by \textsc{SEOBNRv5PHM} drives it, remain uncertain.

\section{Conditioning on a generalized precession parameter}\label{app:chip-gen}

The standard \chip of Eq.~(\ref{eq:chip}) is not conserved under spin precession and heuristically selects the ``dominant'' BH to quantify precession, motivating other and corrected parameterizations. Reference~\cite{Thomas:2020uqj} devises a precession vector that independently treats each BH. Alternatively, Ref.~\cite{Gerosa:2020aiw} notes \chip inconsistently averages over precession cycles for the azimuthal angle between the spin vectors $\phi_{12}$ and proposes a ``generalized \chip'' notion that consistently retains all precession-timescale oscillations:
\begin{equation}\label{eq:chipgen}
\begin{aligned}
    \chipgen &= \bigg[\left(a_1\sin\tau_1\right)^2 + \left(q\frac{3+4q}{4+3q}  a_2\sin\tau_2\right)^2\\
    &+ 2 q\frac{3+4q}{4+3q} a_1a_2\sin\tau_1\sin\tau_2\cos\phi_{12}\bigg]^{1/2}.
\end{aligned}
\end{equation} 
The generalized parameter $\chipgen\in[0,2]$ incorporates contributions from both BHs to spin precession, with the region $\chipgen>1$ restricted to binaries with two precessing spins \cite{DeRenzis:2022vsj}. We test whether \chipgen explains more component spin posterior content than standard \chip and find the two parameters produce nearly-identical constraints for the majority of events.

\begin{figure*}
\centering
\includegraphics[width=0.78\textwidth]{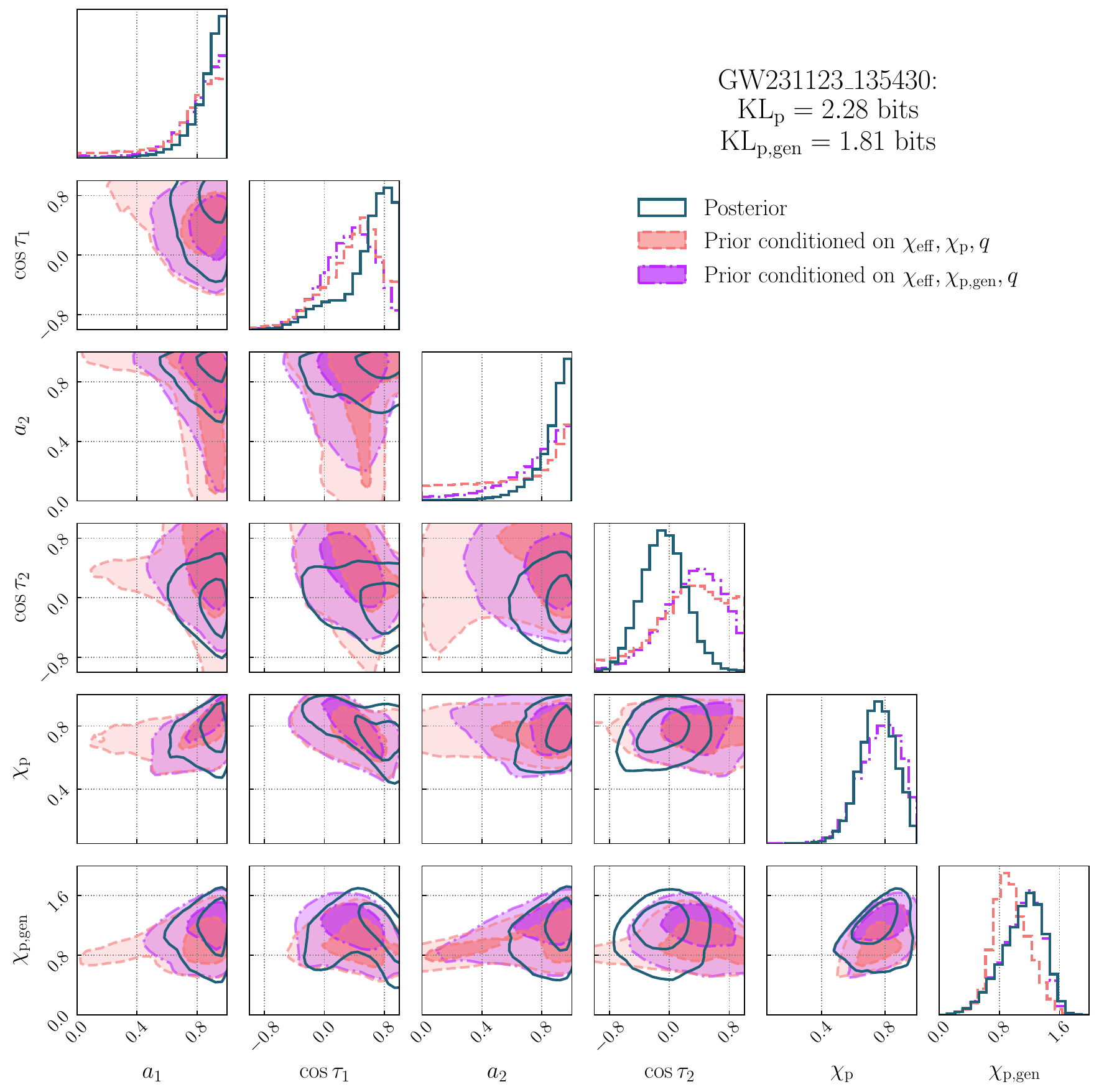}
\caption{Comparison of the posterior (blue) on spin parameters to the $(\chieff,\chip)$-conditioned prior (pink) and the $(\chieff,\chipgen)$-conditioned prior (purple) for GW231123. Contours enclose 50\% and 90\% credible regions.}
\label{fig:spin-corner-gen-231123}
\end{figure*}

The only event with a $>0.1$ bit difference in \spinformation when conditioning on \chip versus \chipgen is GW231123, for which \klp$\approx2.3$ bits while the analogous KL\textsubscript{p,gen}$\approx1.8$ bits, shown in Fig.~\ref{fig:spin-corner-gen-231123}. The difference between its $(\chieff,\chip)$- and $(\chieff,\chipgen)$-conditioned priors is primarily in spin magnitude: \chipgen requires more support at maximal spin magnitudes than \chip and thus better matches the posterior. The support at zero spin magnitude in the $(\chieff,\chip)$-conditioned prior is attributable to the ``max'' operation in \chip, which can allow for more freedom in the spin magnitude of the non-dominant BH than \chipgen. GW231123's $\phi_{12}$ is also one of the best-measured in the catalog, which further contributes to component constraints under \chipgen. Interestingly, \chipgen explains 0.08 bits less of the component spin posterior of GW241011 than \chip. Since \chip and \chipgen are identical in the asymmetric mass limit~\cite{Gerosa:2020aiw}, it is not surprising \chipgen does not perform better than \chip for GW241011.
The catalog results suggest correlations between $\phi_{12}$ and component spins are subdominant, and the influence of combining both BHs' spins is usually small, so \chipgen is not typically more constraining of component spins at current sensitivity.

\end{document}